\documentclass[aps,prb, twocolumn,
reprint,
nofootinbib,
 amsmath,amssymb,
floatfix,
]{revtex4-2}
\usepackage[utf8]{inputenc}
\usepackage{graphicx}
\usepackage{bbold}
\usepackage{bm}
\usepackage{physics}
\usepackage[none]{hyphenat}
\usepackage{hyperref}
\usepackage[shortlabels]{enumitem}
\usepackage{orcidlink}
\usepackage{xcolor}
\hypersetup{
    colorlinks=true,
    linkcolor=black,
    citecolor =blue,
    pdftitle={WeirdFluxModel},
    }
\newcommand{\vect}{\bm}
\newcommand{\mote}{MoTe$_2$ }

\begin{document}

\title{Topological phases of a generalised tripartite Haldane model}

\author{Ahmed Al-kharusi\orcidlink{0009-0006-7981-8498}}\email{ahmed.al-kharusi@postgrad.manchester.ac.uk}\thanks{On leave from the Science Department, Rustaq College of Education,
University of Technology and Applied Sciences, PO Box 10, Al Rustaq 329,
Sultanate of Oman}
\affiliation{Department of Physics and Astronomy, The University of Manchester, Manchester M13 9PL, UK} 

\author{Alessandro Principi\orcidlink{0000-0002-4776-6965}}
\affiliation{Department of Physics and Astronomy, The University of Manchester, Manchester M13 9PL, UK}
\author{Niels R.~Walet\orcidlink{0000-0002-2061-5534}}
\affiliation{Department of Physics and Astronomy, The University of Manchester, Manchester M13 9PL, UK}
\date{\today}
\begin{abstract}
We present a generalised tripartite Haldane model with a complex nearest neighbour hopping parameter. The total magnetic flux through the primitive unit cell, which consists of three hexagonal unit cells is zero, and thus the model has the structure of a loop-current model.  We calculate topological phase diagrams of the Chern numbers as a function of the phases of the nearest and next-nearest neighbour hopping parameters for a fixed ratio of the magnitude of the hopping parameters. We show that, unlike the Haldane model, the topological phase diagram of this model is very complex, but some aspects can still be dealt with analytically. Furthermore, the  Chern numbers of the topological phases are as large as $7$ in absolute value. The analysis is supported by explicit expressions for  energy bands crossing at high symmetry points in the Brillouin zone, which show as linear phase boundaries in the phase diagram. We show that such lines explain many features in the topological phase diagrams. Finally, we analyse a few representative examples of the nature of level crossings away from high symmetry and their evolution with model parameters.    
\end{abstract}

\maketitle  

\section{Introduction}
The Haldane model~\cite{haldane_model_1988} is a foundational example in the understanding of topology in condensed matter physics, since it shows that in two-dimensions it is possible  to have a quantised Hall conductivity without a net magnetic flux through the unit cell. It is described by a tight-binding model for a graphene-like system with a real nearest neighbour hopping parameter and a chiral, and therefore complex, next-nearest neighbour hopping parameter. It is the first model for what we now call a Chern insulator, see, e.g., the review ref.~\cite{liu_recent_2024}. In these materials the Hall conductivity is quantised in the absence of a magnetic field. In particular, the Hall conductivity of a Chern insulator is proportional to a topological invariant known as the Chern number $C$, $C\in\mathbb{Z}$. The quantisation of the Hall conductivity in a Chern insulator is thus analogous to the integer quantum Hall effect~\cite{klitzing_new_1980}, but without the need for of a flux through unit cell.

Building on the work by Haldane, it was predicted~\cite{neupert_fractional_2011, regnault_fractional_2011} that the Hall conductivity for some lattice models can take fractional values, analogous to the fractional quantum Hall effect~\cite{tsui_twodimensional_1982}. Such lattice models are known as fractional Chern insulators~\cite{liu_recent_2024}. Recently, fractional Chern insulating states have been detected experimentally at zero magnetic field in twisted \mote~ \cite{cai_signatures_2023, xu_observation_2023, park_observation_2023, zeng_thermodynamic_2023} and in multilayer graphene systems~\cite{lu_fractional_2024}.

Both Chern insulators and fractional Chern insulators are examples of what are now called topological materials~\cite{ando_topological_2013,hasan_colloquium_2010}, which is one of the most active research fields of condensed matter physics. As a foundational work in this field, many extensions or generalisations of the Haldane model~\cite{haldane_model_1988} have been studied, see, e.g., ~\cite{ikegami_topological_2024,chen_symmetry_2023, sorn_bilayer_2018, nag_extended_2025, wang_higherorder_2021, liu_topological_2012, tupitsyn_phase_2019, lee_interplay_2024, ablowitz_generalized_2024, dabiri_construction_2025, moustaj_latent_2025, mondal_topological_2022, mondal_topological_2021, nag_extended_2025, lahiri_second_2024, yi_higherorder_2023, goncalves_dirac_2019, priestley_haldane_2024, wright_realising_2013, mondal_topological_2023, cheng_topological_2018, parui_topological_2024, andrijauskas_threelevel_2015, bhattacharya_quenching_2017, wang_optically_2012, dey_unconventional_2020, mondal_bandengineered_2023, Sil_2020}.

In this work, we present a tripartite generalised Haldane model. The model is defined on a hexagonal lattice, and  just as in the Haldane model, each hexagon contains two atoms, and the next-nearest-neighbour hopping parameter is complex. The primitive unit cell consists of three hexagons (i.e., 6 atoms). In the Haldane model, the magnetic flux $\Phi$ through any hexagonal unit cell is zero. In our model the total flux through the primitive unit cell is zero, but it is zero in only one of the three hexagonal consitutents of the primitive unit cell, while the other two hexagons have opposite flux, $\Phi=\pm \Phi_0$, where $\Phi_0$ can take any real value. Mathematically, this is captured by having a complex nearest neighbour hopping parameter.\par

There are several reasons to study this tripartite generalised Haldane model. Firstly, it has a simple analytic form, but just too complex for full analytical evaluation. Nevertheless, some very useful analytical results can still be obtained. Secondly, it is just complex enough so that it has a very rich topological phase diagram. In particular, the numerical values of the Chern number $C$ can reach~$\pm7$. Such values are much larger than in, for example, the Haldane model, where $C=0$ or $C=\pm 1$. By utilising these advantages, we study theoretically and numerically the topological phases of the model. Using a fully numerical approach, supported by analytical results, we present an accurate topological phase diagram of the model. We also study typical crossings at the boundaries of the topological phases. We show that at the high-symmetry points it is possible to calculate the energy bands analytically. Further, by equating such expressions for the energy bands, we find analytic expressions for a sizeable fraction of the boundaries of the topological phases.\par

Another motivation to study the generalised tripartite Haldane model is its potential experimental relevance. In particular, the interaction between electrons in a lattice may result in static interatomic currents, forming flux patterns known as loop currents~\cite{fernandes_loopcurrent_2026, chakravarty_hidden_2001, dong_interlayer_2025, dong_loopcurrent_2023, hsu_two_1991, khalifa_lattice_2023, sun_timereversal_2008, varma_nonfermiliquid_1997, venderbos_symmetry_2016, wu_pair_2023, zhu_ordered_2013, dong_interlayer_2025}. Interestingly, the generalised tripartite Haldane model seems similar to an ordered loop current state propsed for graphene in Ref.~\cite{zhu_ordered_2013}, where the loop currents are generated by nearest-neighbour interactions. A flux pattern that is similar to that of the generalised tripartite Haldane model has also been studied in twisted-bilayer graphene~\cite{bultinck_ground_2020}.

This paper is organised as follows. In Sec.~\ref{sec:model}, we introduce the generalised Haldane model. In Sec.~\ref{sec:method}, we discuss the numerical method for calculating the Chern numbers. In Sec.~\ref{sec:symmetries}, we briefly study the symmetries of the model. We present and discuss the results in Sec.~\ref{sec:results}.  Finally, conclusions and possible future work are presented in Sec.~\ref{sec:conclusion}.

\section{Model} \label{sec:model}
\begin{figure}[tbh!]
\includegraphics[width=0.7\linewidth]{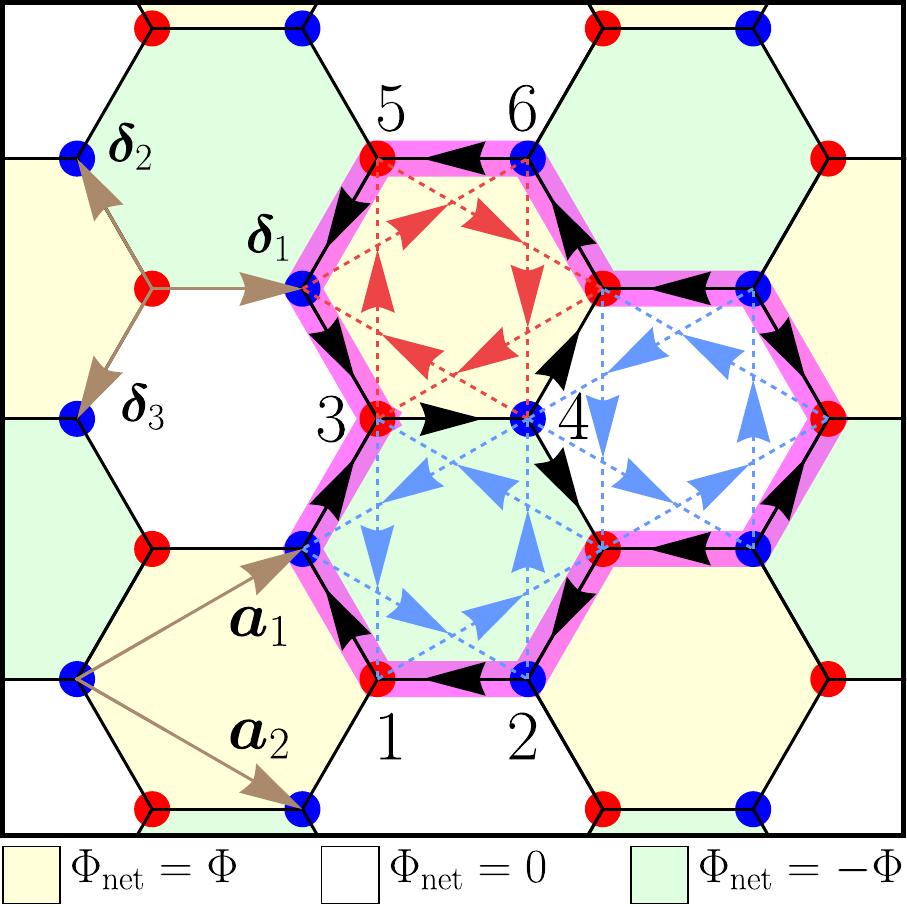}
    \caption{ 
    A graphical representation of the generalised tripartite Haldane model. The sign of the phase of the nearest-neighbour hopping $t_1e^{\pm i \theta}$ is positive if the  hopping is in the direction of the black arrows, and negative otherwise. Similarly, the dotted arrows show the sign of the phase in the next-nearest-neighbour hopping $t_2e^{\pm i \phi}$. The dotted lines are coloured in blue and red to easily distinguish between the clockwise and the anticlockwise directions. The magenta lines show the primitive unit cell, which contains six atoms numbered $1$ to $6$. The graphene unit cells are coloured according to the net magnetic flux. We use light yellow and light green  for $\Phi_\text{net}=\Phi$ or $-\Phi$, respectively, while white indicates where the net flux is zero.}
    \label{fig:HaldaneVsWeirdFluxModel}
\end{figure}

Our model consists of a hexagonal arrangement of atoms in an infinite two-dimensional plane as depicted in Fig.~\ref{fig:HaldaneVsWeirdFluxModel}. The lattice points are generated by the lattice vectors $\bm{a}_1=a(\frac{\sqrt{3}}{2},\,\frac{1}{2})$ and $\bm{a}_2=a(\frac{\sqrt{3}}{2},\,-\frac{1}{2})$, where $a$ is the lattice constant. A primitive unit cell of this periodic structure contains two atoms, which are indicated by the red and blue dots in Fig.~\ref{fig:HaldaneVsWeirdFluxModel}. The position of a red atom is related to that of a blue atom by one of the three vectors $\bm{\delta}_1=a(\frac{1}{\sqrt{3}},\,0)$, $\bm{\delta}_2=a(-\frac{1}{2\sqrt{3}},\,\frac{1}{2})$ and $\bm{\delta}_3=a(-\frac{1}{2\sqrt{3}},\,-\frac{1}{2})$ (see Fig.~\ref{fig:HaldaneVsWeirdFluxModel}).

We can obtain the Haldane model by assigning an on-site energy $\pm M$ to the blue (red) atoms, have a real nearest-neighbour hopping parameter $t_1$ and a (chiral) complex next nearest-neighbour hopping parameter $t_2e^{\pm i\phi}$, where $t_2$ and $\phi$ are real numbers, and the sign in the exponent of the hopping parameter is opposite for left and right-handed hopping. The phase $\phi$ can, for instance, be conceptualised as caused by a localised magnetic field that is perpendicular to the lattice plane. The flux of this magnetic field through the primitive unit cell is zero; consequently, no phase is acquired when hopping to nearest neighbours. By changing the parameters $(M,\,t_2/t_1,\,\phi)$ of this model, one can control the topology of the system. In particular, the two energy bands of the Haldane model can be topological, in which case their Chern numbers are $\pm 1$, or they can be trivial, in which case their Chern numbers are zero. 

We modify Haldane's tight-binding~\cite{haldane_model_1988} model by allowing the nearest-neighbour hopping parameter $t_1$ to be also complex, and depend on the bond under consideration. We write $t_1\rightarrow t_1 e^{\pm i\theta}$, where $t_1$ and $\theta$ are real numbers. We shall use $M=0$. To define the phases used for each bond, consider the yellow hexagonal unit cells in Fig.~\ref{fig:HaldaneVsWeirdFluxModel}. For atoms on the edge of these unit cells, we choose the positive sign of the phase if the nearest neighbour hopping is in the anticlockwise direction, and negative if clockwise. Since the green cells share links with yellow ones, and those links have been allocated a phase, in that cell we choose a positive sign in the exponent if the nearest neighbour hopping is in the clockwise direction. These choices fix the nearest neighbour hopping parameters in the white unit cells. This lack of freedom is one of the sources of geometric frustration. The primitive unit cell of this tight-binding model contains one yellow, one green and one white hexagon, containing six atoms, as indicated by the magenta lines in Fig.~\ref{fig:HaldaneVsWeirdFluxModel}. Use of the Peierls substitution and summing these phases along closed loops implies that the net flux $\Phi_{\text{net}}$ through the yellow, green and white cells is $\Phi$, $-\Phi$ and $0$, respectively. This depends on $\theta$ as  $\Phi=6\theta \,\Phi_0$, where $\Phi_0=h/e$ is the quantum of flux.

Following the Haldane model, we introduce a complex nearest-neighbour hopping parameter $ t_2 e^{\pm i\phi}$, where $t_2$ and $\phi$ are real numbers. There is a freedom in choosing the sign of the phases. In particular, one can choose the sign of the phase to be positive if: (i)~the next-nearest-neighbour hopping is in the clockwise direction in the yellow unit cells and anticlockwise in the green one--consistent with the choice of nearest-neighbour phase. We can then choose the direction in the white cell using either recipe, leading to equivalent models; (ii)~the positive phase of the next-nearest-neighbour hopping is in the clockwise direction in the yellow and green unit cells and anticlockwise in the white unit cells; (iii)~the next-nearest-neighbour hopping is the clockwise direction in all unit cells. 

In this paper, we fix the sign of $\phi$ according to rule (i), as depicted in Fig.~\ref{fig:HaldaneVsWeirdFluxModel}, with the white cell following the green one, which leads to a $6\times6$ tight-binding Hamiltonian, which is presented in detail in Appendix~\ref{sec:AppendixA}.  Due to zone folding with three hexagonal cells we have a smaller first Brillouin zone, with high symmetry points labelled as $\gamma=(0,\,0)$, $k_-~=~\frac{2\pi}{3a}(\frac{1}{\sqrt{3}},\,1)$, $k_+~=~\frac{2\pi}{3a}(-\frac{1}{\sqrt{3}},\,1)$, $k_+^{\prime}=-\frac{2\pi}{3a}(\frac{1}{\sqrt{3}},\,1)$ and $k_-^{\prime}~=~\frac{2\pi}{3a}(\frac{1}{\sqrt{3}},\,-1)$. To highlight that the first Brillouin zone is smaller than that of the standard Haldane model, we used lowercase letters to label the high-symmetry points. Note that by definition of the Brillouin zone, the energy bands at $k_{\pm}$ are equivalent to the energy bands at $k^{\prime}_{\pm}$, respectively. Hence, hereafter, we will discuss the results for $k_{\pm}$, ignoring $k^{\prime}_{\pm}$.

\section{Method}\label{sec:method}
Since there are $6$ atoms per unit cell, solving the tight-binding model is equivalent to 
diagonalising a $6\times 6$ matrix. This gives us the band structure and eigenvectors. We then calculate the Chern numbers by integrating over $k$.
However, before calculating the Chern number of a band, we must ensure that it is (well) separated from the other ones. As an operational definition, for given values of $(t_1,\,t_2,\,\theta,\,\phi)$, we consider a band to be well separated from the rest if the energy difference between the band of interest and any other satisfies $\Delta E\geq 10^{-4}t_1$. 
This finite limit avoids having to integrate over a very fine mesh to capture an extremely localised Berry curvature.
This reflects the fact that the Chern number is only well defined for an isolated band, i.e., $\Delta E>0$. In Fig.~\ref{fig:Ct2byt1is4by10}, we found that the ratio of data points where $\Delta E<10^{-4}t_1$ to the total number of data points in the phase diagrams is less than $2\%$. The large majority of those cases is on a crossing line.

We then calculate the Chern number for each band by integrating the Berry curvature over the full Brillouin zone. In particular, the Chern number for band $n$ is given by~\cite{xiao_berry_2010} 
\begin{equation}\label{eq:ChernNo}
    C_n = \frac{1}{2\pi} \int_{\in\text{FBZ}} d\vect{k}\,\Omega_n(\vect{k}),
\end{equation}
where the integral is over the first Brillouin zone. The Berry curvature $\Omega_n(\vect{k})$ for band $n$ is calculated as
\begin{equation}\label{eq:BerryCurvature}
\Omega_n(\vect{k}) =-2 \Im \sum_{n^\prime\neq n}\frac{\matrixel{u_{n\vect{k}}}{\partial_x H}{u_{n^\prime\vect{k}}}\matrixel{u_{n^\prime\vect{k}}}{\partial_y H}{u_{n\vect{k}}}}{(E_n-E_{n^\prime})^2},
\end{equation}
where $\{{u_{n\vect{k}}}\}$ and $\{E_n\}$ are the eigenstates and energies of the tight-binding Hamiltonian $H$ and 
$\partial_{x(y)}\equiv \partial_{k_{x(y)}}$.
The derivatives of the Hamiltonian $\partial_x H$ and $\partial_y H$ can be calculated analytically~(see Appendix~\ref{sec:AppendixA} for the analytic expression of $H$). 

At each $\vect{k}$-point in the BZ, the Hamiltonian is evaluated for a given set of parameters ($t_1$, $t_2$, $\theta$, $\phi$) and diagonalised numerically to find 
$\{{u_{n\vect{k}}}\}$ and $\{E_n\}$. Using these results, Eq.~(\ref{eq:BerryCurvature}) is evaluated. Further, by numerically integrating $\Omega_n(\vect{k})$ over the BZ as given by Eq.~(\ref{eq:ChernNo}), the Chern number can be obtained.

\section{Symmetries}\label{sec:symmetries}
The tight-binding Hamiltonian of the generalised Haldane model $H_{\bm{k}}$ has some useful symmetries under translation of $\theta$ and $\phi$. In particular, $H_{\bm{k}}$ satisfies
\begin{align}\label{eq:Hsymm1}
   H_{\bm{k}}(t_1,\,t_2,\,\theta+\pi,\,\phi)&= U^{\dagger} H_{\bm{k}}(t_1,\,t_2,\,\theta,\,\phi) U, \\
\label{eq:Hsymm2}
  H_{\bm{k}}(t_1,\,t_2,\,\theta,\,\phi+\pi)&= -U^{\dagger}  H_{\bm{k}}(t_1,\,t_2,\,\theta,\,\phi) U,    
\end{align}
where $U=\operatorname{diag}(i,\,-i,\,i,\,-i,\,i,\,-i)$. Eq.~(\ref{eq:Hsymm1}) implies that the Chern number is invariant under $\theta\rightarrow\theta+\pi$. Since the minus sign in Eq.~(\ref{eq:Hsymm2}) implies that the ordering of the bands inverts, it links the topology of top and bottom bands, etc.
Finally, the transformation $(\theta,\,\phi)\rightarrow(-\theta,\,-\phi)$ reverses the sign of the Berry phase (see Fig.~\ref{fig:HaldaneVsWeirdFluxModel}), see Appendix~\ref{sec:AppendixD}. Thus, under such a transformation, the sign of the Chern number also changes $C\rightarrow-C$. 

This set of three symmetries show that the full topological phase diagrams for all energy bands of the generalised Haldane model can be generated from the topological phase diagrams of just the first three energy bands in the range $\theta\in(0,\,\pi]$ and $\phi\in(0,\,\pi]$. 

Finally, if we split $H$ in a nearest neighbour and next-nearest neighbour part, we can easily show that these two commute at the high symmetry points, 
thus showing that the eigenvalues must be linear in both $t_1$ and $t_2$.
They have
simple and analytical solutions 
as shown in appendix \ref{sec:AppendixB}.
The structure of (some of) the eigenvalues between the $k$ and $\gamma$ points shows some similarities, which might be suggestive of other closed form solutions. However, as far as we have been able to determine, no other points with generic closed form solutions exist.

\section{Results and discussion} \label{sec:results}

\begin{figure}[tb]
\includegraphics[width=0.7\linewidth]{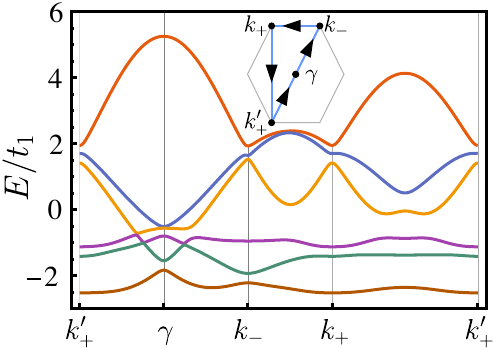} 
    \caption{The energy bands of the generalised Haldane model for $t_2/t_1=4/10$, $\theta=\pi/12$ and $\phi=\pi/15$. There is no actual band crossing in this plot, but  the apparent crossing that can be seen shows the narrowness of the band gaps. The inset shows the path taken through the BZ connecting the high symmetry points. }
    \label{fig:bandt1byt2is4by10}
\end{figure}

\begin{figure}[tbh!]
\includegraphics[width=0.95\linewidth]{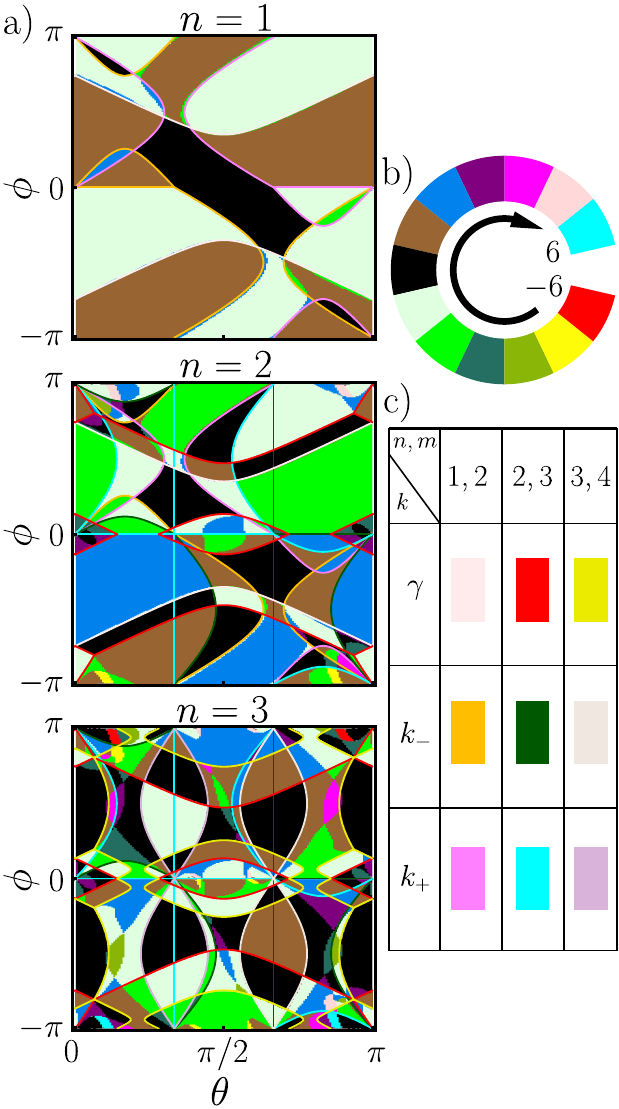}
    \caption{(a) The topological phase diagrams of the upper three bands of the generalised Haldane model, where $n$ is the band index. The bands are numbered from highest energy value to lowest energy value such that $n=1$ is the uppermost energy band, and $n=6$ is the lowermost energy band. The phase diagrams show the Chern number as a function of the phases $\theta$ and $\phi$, where $t_2/t_1=4/10$. The numerical calculations were done only for $\phi\in[0,\,\pi ]$, and the other halves of the phase diagrams, namely for $\phi\in[-\pi,\,0 )$ were generated by symmetry as explained in the main text.  Calculation were performed on a grid of $(\theta,\,\phi)$ values of size  $N_{\theta}\times N_{\phi}=161\times 321$. Each point is then coloured by its Chern number, while data points where the calculated Chern number is not an integer within a precision of $10^{-4}$ are coloured gray. Data points where the value of the band gap is less than $10^{-4} t_1$ are coloured white. (b) Colour coding of the Chern numbers used in (a), in a range from $-6$ to $6$. (c) Colours used for lines showing band-crossing at a high-symmetry point in (a): we colour both by point and the crossing bands. These are based on the analytical expression for the crossing lines  in Appendix~\ref{sec:AppendixB}.}
    \label{fig:Ct2byt1is4by10}
\end{figure}

We now present the topological phase diagram of the generalised Haldane model, highlighting its complexity.  
For that reason we concentrate on a single case of the ratio of hopping amplitudes,  $t_2/t_1=4/10$. We select this ratio, since there is a large number of crossings which is particularly useful for the discussion of the nature of topological phase transitions. The complexity of the topological phase diagrams for other ratios of $t_2/t_1$ are discussed in Appendix~\ref{sec:AppendixC}.  We concentrate on the top three bands, using the symmetries discussed above. 
An example of the energy bands for $(\theta,\,\phi)=(\pi/12,\,\pi/15)$ is presented in Fig.~\ref{fig:bandt1byt2is4by10}. This is an interesting and not atypical case. Although  there are no real band crossings, there are several points where the energy bands are nearly degenerate. If we are to understand the topology through the calculation of  Chern numbers, this shows that we have to be careful since rather large contributions are found near every narrowly avoided crossing.

Since we can use the symmetries discussed in Sec.~\ref{sec:symmetries} to relate the Chern number in the upper three bands to the Chern number of the lower three by the transformation $(\theta,\phi)\rightarrow(\pi-\theta,\pi-\phi)$, $n\rightarrow 7-n$, which simply inverts the Chern numbers, we only need to concentrate on the upper three.
The topological phase diagrams, the Chern number as a function of  $\theta$ and $\phi$, for those bands are depicted in Fig.~\ref{fig:Ct2byt1is4by10}(a).

\begin{figure}[!tb]
\includegraphics[width=\linewidth]{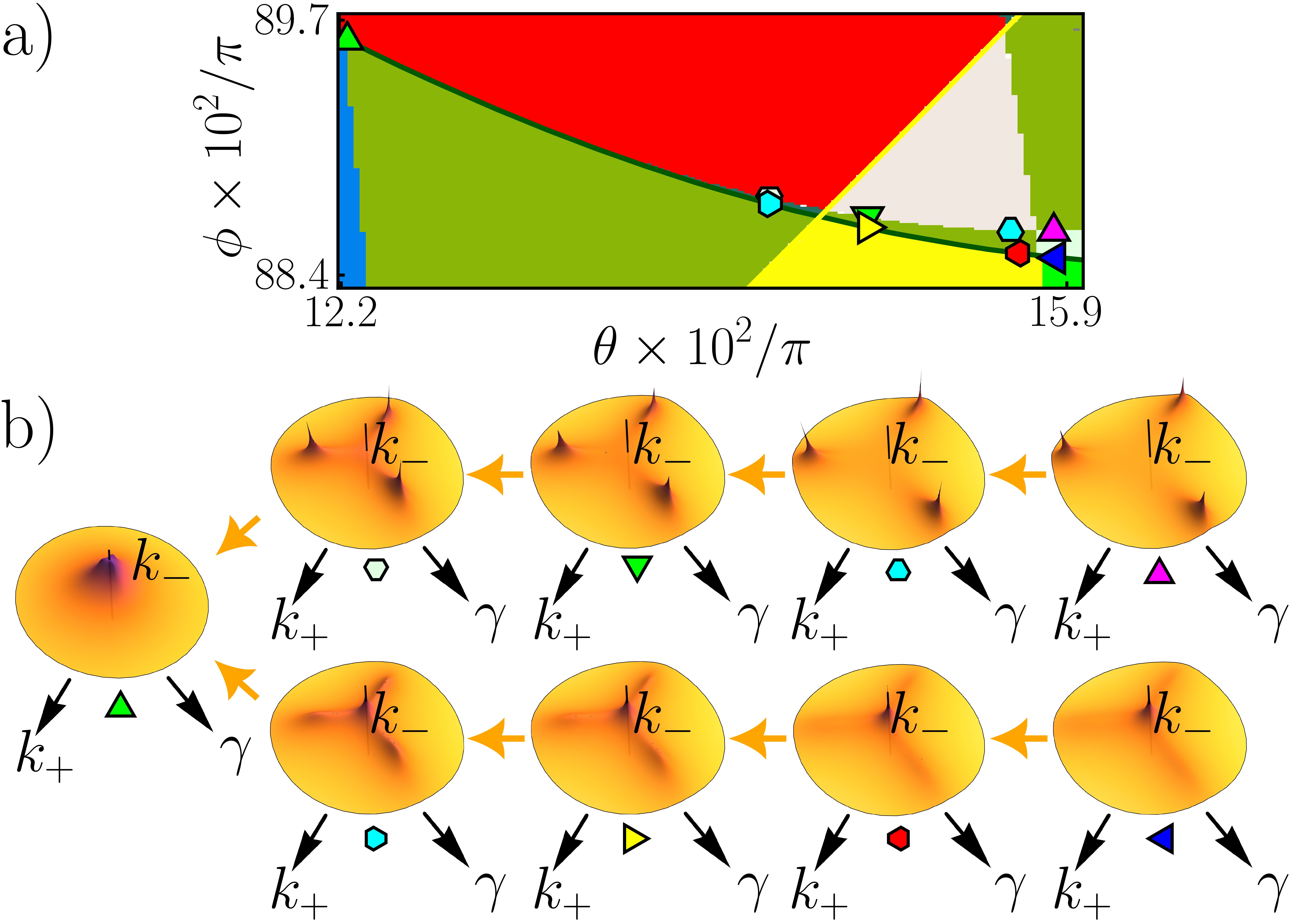}
    \caption{An example of a merger between two different types of crossings. (a) a zoomed-in plot of the topological phase diagram of the third energy band; colour coding as in Fig.~\ref{fig:Ct2byt1is4by10}. In the light brown area, the Chern number is equal to $-7$.  Markers are plotted in (a) along two different crossing lines. The dark green line (see Fig.~\ref{fig:Ct2byt1is4by10}(c)) is associated with a crossing at $k_-$, while the other line is associated with three crossings $\{k_i\}$, away from the high symmetry point. (b) Plots of $-\ln(E_2-E_3)$ for the two crossing lines in the vicinity of $k_-$. Each peak in~(b) indicates a crossing between $E_2$ and $E_3$. The radii of the plots in (b) are equal to $k/a=0.3$. See main text for discussion. }
    \label{fig:MergingCrossingLines}
\end{figure}

As can be seen in Fig.~\ref{fig:Ct2byt1is4by10}(a), the complexity of the phase diagram increases as the band index increases from $1$ to $3$. For example, the number of distinct topological phases for $n=3$ is much larger than that for $n=1$. Additionally, the minimum and maximum Chern numbers for $n=1$ and $n=3$ are $\pm 2$ and $\pm 7$, respectively. If the resolution of the phase diagram for $n=3$  is increased, we find very small regions where the Chern number is even larger in absolute value. For example, in Fig.~\ref{fig:MergingCrossingLines}(a), which shows a zoomed-in plot of the topological phase diagram for $n=3$~(see Fig.~\ref{fig:Ct2byt1is4by10}), there is a region where the Chern number is $-7$. Such a region is not visible in Fig.~\ref{fig:Ct2byt1is4by10}. Since finding the maximum or minimum Chern numbers is not the primary goal in this work, we will focus instead on the complexity and the general structure of the topological phase diagrams. 

At the boundary between any pair of distinct topological phases of the band $n$, $E_n$ must be degenerate with another energy band $E_m$. 
In general it does not seem to be possible to find analytic expressions for the $k$ points where this crossing occurs. However, if we concentrate on crossings at high-symmetry points  in the BZ, i.e., $k=\gamma,\,k_-$ or $k_+$, analytic expressions can be derived for the lines in the $\theta-\phi$ plane where such crossings occur. We will refer to such lines as the analytic crossing lines;  their analytical expressions are presented in Appendix~\ref{sec:AppendixB}. As can be seen in  Fig.~\ref{fig:Ct2byt1is4by10}(a), the analytic crossing lines explain a substantial fraction of the topological phase transitions. The intersection of any two of the high-symmetry-point crossing lines is rather trivial: we just have Dirac cones at two isolated points in the BZ. If we follow along any of the crossing lines, points near the intersection only have a single crossing. 

Looking at the figure in a bit more detail, we see that the change in the Chern number between regions in  Fig.~\ref{fig:Ct2byt1is4by10}(a) that are separated by the analytic crossing lines is normally $1$, indicating that the crossing is linear in $k$: at the crossing we obtain a singularity  in the Berry phase, where we get opposite $\pm1/2$ fluxes in the top and bottom band, corresponding to a $\pm 1$ monopole. This implies that the difference in Chern number on both sides of the line is one. As we shall see, this is fully consistent with our analysis below.

The situation looks to be more complex for the other crossing lines, where we usually we see a change by $3$ or $6$ units.
Deeper investigation of the band structure shows that these are still isolated crossings. However, due to symmetry and features such as  trigonal warping, there are multiple (typically $3$ or $6$) isolated crossing points.  Finally, at a few isolated points we see quadratic crossing: these are extremely rare.
Of more interest is the typical nature of the crossings of bands away from the high symmetry points, and especially the intersection of band-crossing lines.  

\begin{figure}[!tb]
\includegraphics[width=\linewidth]{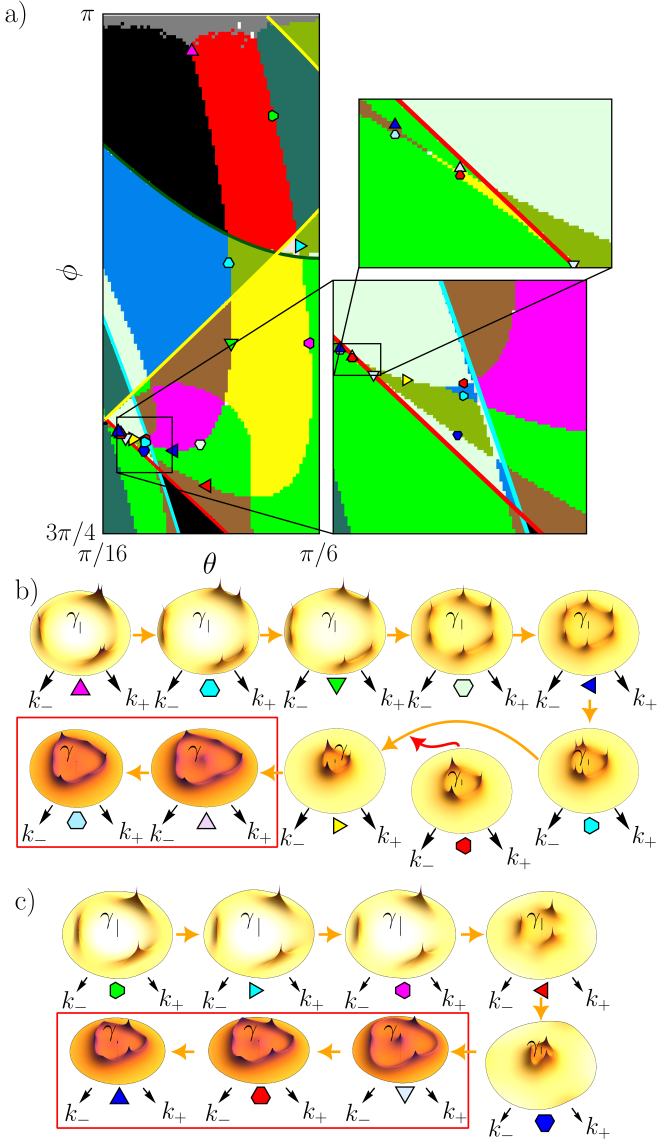}
    \caption{Examples of lines in the phase diagrams where the second and third energy bands cross at a set of points $\{k_i\}$, where none of $\{k_i\}$ is a high symmetry point. (a) A zoomed-in region of the topological phase diagram of the third band; colour coding as in Fig.~\ref{fig:Ct2byt1is4by10}. Markers are plotted along two different separatrices in the phase diagram, and each marker in a) corresponds to the plot with the same marker in b) and c). The plots in (b) and (c) show $-\ln(E_2-E_3)$ in the vicinity of $\gamma$ in the BZ. Thus, each peak indicates a crossing between the second and third energy bands. The radius of the disks inside the red rectangles in (b) and (c) is $k/a=0.3$, while the radius of the other plots in (b) is $k/a=1$ and (c) $k/a=3/2$.  See main text for discussion.}
    \label{fig:nonAnaltCrossLns}
\end{figure}

So let us study a few typical examples that support our assertions. First of all, 
in Fig.~\ref{fig:nonAnaltCrossLns} we analyse two examples of lines in the phase diagram of the third band where the crossing occurs away from the high-symmetry points in the BZ. This is based on a detail of the topological phase diagram of the third band in Fig.~\ref{fig:Ct2byt1is4by10}(a), where we label points of interest by markers. Fig.~\ref{fig:nonAnaltCrossLns}(b) and Fig.~\ref{fig:nonAnaltCrossLns}(c) show plots of the spectral gap--since we are interested in where it vanishes we plot $-\log(\Delta E)$--in $k$ space near the $\gamma$ point, along the left and right lines in Fig.~\ref{fig:nonAnaltCrossLns}(a). The markers near the gaps refer to the same marker in part a) of the figure.

We follow those lines as they emerge from (or cross at)  $\theta=\pi$, to the apparent terminal point which occurs in a very busy area in the lower corner of the figure, see the insets for additional details.
Across the left-most line, we initially find a change in Chern number of six units. This is easily explained from the plots of the gap,
where we seem to start with a double (quadratic) crossing near $\phi=\pi$ along the $\gamma k_+$ line, which develops into six isolated crossing that move to the $\gamma-m$ lines, where $m$ is the midpoint between $k_{+}$ and $k_{-}$. As the line curves to the lower left, these seem to coincide again, but now along the $\gamma k_-$ line. At that point the line labelled by the red hexagon merges. It carries a charge (change in Chern number) of three, and annihilates with half of the crossings on our initial lines, which now continues with a charge of three. Initially, this seems to end at the high symmetry line (red line, $\gamma$ point crossings). 
That could be possible if the second line that we will study below carries the opposite charge (which it does). However, as we zoom in even further, we see this is not the case, and the situation is even more complex, and the line crosses the high symmetry line.

Before continuing this analysis, let us look at the rightmost line--selected since it seemed to end at the same point as the first line.
As depicted in Fig.~\ref{fig:nonAnaltCrossLns}(c), the second line has three crossings, and the associated change in the Chern number is indeed three. The three crossing points are all along the $\gamma k_+$ line, and as we travel along the line the only feature is that these move ever closer to the $\gamma$ point, without ever reaching it. As we can see in the final zoom, it crosses the high-symmetry line, and then crosses with the first line as well, continuing beyond that. From the change in Chern numbers we see these two cases have opposite charges. However, they do not annihilate since the crossing points occur along either $\gamma k_-$ or $\gamma k_+$.

In Fig.~\ref{fig:MergingCrossingLines} we show an example of a crossing line merging with a high-symmetry crossing, so a combination of charges three and minus one. Again, Fig.~\ref{fig:MergingCrossingLines}(a) is a zoomed-in detail of the topological phase diagram of the third energy band in Fig.~\ref{fig:Ct2byt1is4by10}(a). Fig.~\ref{fig:MergingCrossingLines}(b) shows plots of the logarithm of the band gap between the second and third energy bands at different points along the two merging lines. As can be seen in Fig.~\ref{fig:MergingCrossingLines}(b), along one line, the crossing is at $k_{-}$, while along the other line, there are triple crossing points in the BZ near $k_{-}$. As we move along the line, these triple crossing points move inward towards $k_-$. When the two types of crossings merge,
we get an apparent charge 2 crossing. However, initially this contains four crossings with a net charge of two. As we see from the final point along the line, in the end 
a quadratic crossing appears at $k_{-}$. Clearly a charge two monopole is rather unstable, and we only find this at a single point. The change in the Chern number between regions that are separated by this quadratic crossing is two, as can be seen in Fig.~\ref{fig:MergingCrossingLines}(a) and Fig.~\ref{fig:MergingCrossingLines}(b). 

We have thus seen both the simplicity and complexity of crossing lines. It seems that we have two generic behaviours: the most common one is simply that domain boundaries cross without interacting. The other is the merging of boundaries, where some of the charges annihilate. Due to the three-fold symmetry of the model, trigonal warping is a common occurrence, which means crossing come in charge one (high symmetry) three (along a line connecting high symmetry points) or six (generic crossing point). Only in rare occasions do we see charge 2 quadratic crossings. Even though we can in principle combine all of the above to produce complex states with isolated charges (such as $1+1-3$) these do not seem to be common if they occur at all.

\section{Conclusion} \label{sec:conclusion}
We have presented a detailed analysis of  a generalised tripartite Haldane model. The complexity and fine detail in the phase diagram is surprising for such a simple model, and we need both analytical and  numerical investigations to understand all the details of the phase diagram. In the end the principles are relatively straightforward. 
 We showed that the isolated band crossings are almost always associated with a unit change in the Chern number, as appropriate to a linear crossing. 
Due to threefold symmetry, we see many topological charge changes as a multiple of three when crossings occur away from high symmetry points, and come as three (or six) isolated crossing. The analytic nature of the spectrum at  high-symmetry points, allows us, by requiring a degenerate spectrum, to find analytical expressions for some of the boundaries in the topological phase diagrams. They allow us to show how the typical crossings appear, evolve and merge in the topological phase diagrams.
We also presented more complex behaviours. For instance, crossings at non-high-symmetry points may merge with a crossing at a high-symmetry point. In such a case, the change in the Chern number will be dictated by the boundary line resulting from the merger of the different types of crossings. For example, if a linear crossing at a high-symmetry point merges with a triple linear crossing with an opposite sign of the Berry curvature, then such a boundary line will be associated with a change of two units in the Chern number.

The analogy between the model studied here and the description of loop-current states, such as that in graphene~\cite{zhu_ordered_2013},  suggests also  that the the topology of such states may well be highly complex, an issue that has not yet been studied in detail. There is no reason the methods of this work could not be applied to model using a non-interacting tight binding model the nature of loop-current induced flux states in the more commonly studied Kagome lattice. This is of particular interest since metals with Kagome structure are promising platforms for the possible experimental realisation of loop current states~\cite{fernandes_loopcurrent_2026}.

Furthermore, building on this work, by tuning the ratio of the hopping parameters~\cite{neupert_fractional_2011, regnault_fractional_2011}, one can also search for topological flat bands. Such bands may host exotic topological phases, including non-Abelian topological phases~\cite{reddy_nonabelian_2024,sterdyniak_series_2013}. The existence of such phases can be confirmed by performing exact diagonalisation or many-body techniques such as the infinite density matrix renormalisation group~\cite{grushin_characterization_2015, chen_simulating_2026}.
\section*{Acknowledgment}
The authors would like to acknowledge the assistance given by Research IT and the use of the Computational Shared Facility at The University of Manchester. AA acknowledges support  by the Ministry of Education of Oman. NRW is  supported by the UK Science and Technology Funding Council [grant number ST/Y000323/1].
\par
\bibliography{MyLibraryNewFinal}
\appendix
\section{Matrix elements of the tight-binding model }\label{sec:AppendixA}
The primitive unit cell of the generalised Haldane model has six atoms as depicted in Fig.~\ref{fig:HaldaneVsWeirdFluxModel}. Each atom has three nearest neighbours and six next nearest neighbours. The nearest and next nearest neighbour hopping parameters are complex and can be written as $t_1 e^{i \theta}$ and $t_2 e^{i \phi}$, respectively. The signs of the phases $\theta$ and $\phi$ are determined depending on whether the hopping is in the direction of the arrows which are plotted in Fig.~\ref{fig:HaldaneVsWeirdFluxModel}. In particular, the sign of $\theta$ or $\phi$ is positive if the hopping is in the direction of one of the arrows and negative otherwise. For example, using the convention above, the corresponding matrix elements of the tight-binding Hamiltonian that describe hoppings from atom number 1 to atom number 2 is $H_{12} =   t_1 e^{-i (\theta -\vect{\delta}_1 \cdot\vect{k})}$. Similarly, all matrix elements of the tight-binding Hamiltonian can be deduced,
\begin{widetext}
\begin{equation}\label{eq:fullTBhamiltonain}
\begin{aligned}
H_{12} &=   t_1 e^{-i \theta}e^{i\vect{\delta}_1 \cdot\vect{k}},\\
H_{13} &= t_2 e^{-i\vect{a}_1\cdot\vect{k}} \left(e^{-i \phi}\left(e^{(i (2 \vect{a}_1-\vect{a}_2))\cdot\vect{k}}+e^{(i (\vect{a}_1+\vect{a}_2))\cdot\vect{k}}\right)+e^{ i \phi }\right),\\
H_{14} &=t_1 e^{-i \theta} e^{i\vect{\delta}_3\cdot\vect{k}}, \\
H_{15} &=t_2 e^{-i \vect{a}_2\cdot\vect{k}} \left(e^{-i\phi}+ e^{ i \phi } \left(e^{-i (\vect{a}_1-2 \vect{a}_2)\cdot\vect{k}}+e^{i (\vect{a}_1+\vect{a}_2)\cdot\vect{k}}\right)\right), \\
H_{16} &= t_1 e^{i \theta}e^{i\vect{\delta}_2\cdot\vect{k}} ,\\
H_{23} &=  t_1 e^{i \theta}e^{ -i\vect{\delta}_2\cdot\vect{k}}, \\
H_{24} &=   t_2 e^{-i\vect{a}_1\cdot\vect{k}} \left(e^{i \phi } \left(1+e^{i (2 \vect{a}_1-\vect{a}_2)\cdot\vect{k}}\right)+e^{-i\phi}e^{i (\vect{a}_1+\vect{a}_2)\cdot\vect{k}}\right)  ,\\
H_{25} &=   t_1 e^{-i \theta}e^{-i\vect{\delta}_3\cdot\vect{k}}, \\
H_{26} &=   t_2 e^{-i\vect{a}_2\cdot\vect{k}} \left(e^{i\phi} e^{i (\vect{a}_1+\vect{a}_2)\cdot\vect{k} }+e^{-i\phi} (e^{(-i (\vect{a}_1-2 \vect{a}_2))\cdot\vect{k}}+1)\right), \\
H_{34} &=  t_1 e^{i \theta}e^{i\vect{\delta}_1 \cdot\vect{k}} ,  \\
H_{35} &=   t_2 e^{-i \vect{a}_1\cdot\vect{k}} \left(e^{i\phi}e^{i (2 \vect{a}_1-\vect{a}_2)\cdot\vect{k} }+e^{-i\phi}(e^{i (\vect{a}_1+\vect{a}_2)\cdot\vect{k}}+1)\right), \\
H_{36} &=   t_1 e^{-i \theta}e^{i\vect{\delta}_3\cdot\vect{k}}, \\
H_{45} &=   t_1 e^{i \theta}e^{-i\vect{\delta}_2\cdot\vect{k}},  \\
H_{46} &=    t_2  e^{-i \vect{a}_1\cdot\vect{k}} \left(e^{i\phi }\left(e^{i (\vect{a}_1+\vect{a}_2)\cdot\vect{k} }+1\right)+e^{-i\phi }e^{i (2 \vect{a}_1-\vect{a}_2)\cdot\vect{k}}\right),\\
H_{56} &=    t_1 e^{-i \theta}e^{i\vect{\delta}_1 \cdot\vect{k}}.\\
\end{aligned}
\end{equation}
\end{widetext}

\section{Crossings at high symmetry points}\label{sec:AppendixB}
Let $\{f(\theta,\,\phi,\bm{k}\,,n,\,m)\}$ be a set of lines that separate the different phases in the topological phase diagrams, where $n$ and $m$ label the degenerate energy  bands. For a topological phase transition to occur, two or more energy bands must be degenerate at such lines.\par
In this section, we derive analytic expressions for $\{f(\theta,\,\phi,\bm{k}\,,n,\,m)\}$, where $\bm{k}$ is a high symmetry point, namely $\gamma$, $k_{-}$ or $k_{+}$. To achieve this, we first find analytic expressions for the energy bands $E_n$ at the high symmetry points.   \par
For simplicity, let $t_1=1$ and denote $t_2$ by $\alpha$. The energy bands $E_n$ of the tight-binding Hamiltonian can be found by finding solutions to the following equation
\begin{equation}
\det(H(\bm{k}\,,\,\alpha \,,\theta,\,\phi)-E_n \,\mathbb{1})=0, 
\end{equation}
where $\mathbb{1}$ is the identity matrix and its dimensions is equivalent to $H$ i.e.~$6\times 6$. Let $p(E_n)=\det(H-E_n \,\mathbb{1})$. The polynomial $p(E_n)$ is $6^{\text{th}}$ order in $E_n$ and there are no general solutions to $p(E_n)=0$. However, if $\bm{k}$ is a high symmetry point, namely $\gamma$, $k_{-}$ or $k_{+}$, it is possible to find analytic solutions to $p(E_n)=0$, as we will discuss below.

\subsection{Crossings at the $\gamma$ point}
To simplify the analytic expressions, we make the substitution $\mu_n=E_n+3 \alpha\,\cos(\phi)$.  The six energy eigenvalues $\mu_n$ at $\vect{k}=\gamma$ split in a  set of four and a set of two, with rather different behaviour, which as stated before, are linear in $\alpha$ since the nearest and next-nearest contributions to the Hamiltonian matrix commute at $\gamma$,
\begin{equation}\label{eq:EnAtGamma}
\begin{aligned}
\mu^1_{\sigma_1\sigma_2}&=\sigma_1 2 \sin (\theta )+\sigma_2 \sqrt{3} \alpha  \sin (\phi ),\\
\mu^{2}_{\sigma}&=\sigma\sqrt{5+4 \cos (2 \theta )}+9 \alpha  \cos (\phi ).
\end{aligned}
\end{equation}
Here all the $\sigma$s are $\pm1$.

Our aim now is to find analytic expression for the intersection lines. It is easiest to express those in terms of the complete but unordered $\mu$s. This can then numerically be separated in pieces for each energy band, by ordering the $\mu$s for each value of $\theta,\phi$.
We now introduce a few new functions in order to write the results in a compact form. Let 
\begin{equation}
 r_{\sigma,\sigma_1,\sigma_2,\sigma'}(\theta)= 2\arctan\left(\frac{-\sigma_2\,\sqrt{3} \alpha  +\sigma' \sqrt{84 \alpha ^2-c_{\sigma\sigma_1}^2}}{9 \alpha + c_{\sigma\sigma_1} }\right),   
\end{equation}
 where $c_{\sigma\sigma_1}=-\sigma\sqrt{5+4 \cos (2 \theta )} +\sigma_1 2 \sin (\theta )$.
  All variables $\sigma_i$ can either be $-1$ or $+1$. Further, let 
 \begin{equation}
 q_{\sigma_1\sigma_2}(\phi)= \arcsin\left(\frac{\sigma_2-\sigma_1}{4} \sqrt{3} \alpha  \sin (\phi )\right).
 \end{equation}
 The crossings between the energy bands~(Eq.~(\ref{eq:EnAtGamma})), or the analytic crossing lines at the gamma point, are summarised in Table~I. In all tables of Appendix~\ref{sec:AppendixB}, $n$ is any integer, $n\in\mathbb{Z}$.
\begin{table}[!htb]
    \caption{\label{ee}The analytical crossing lines at the gamma point as a function of $\theta$ or $\phi$. This uses the compact expressions in the main text, and label the intersection with the unordered eigenvalues $\mu^{1,2}$.   Note that the pair of bands $\mu^{2}_{\pm}$ do not cross. }
\begin{tabular}{c|c}
    \hline
      & $\mu^{2}_{\sigma}$  \\
    \hline 
        $\mu^{1}_{\sigma_1\sigma_2}$ & $\phi=r_{\sigma,\sigma_1,\sigma_2,\pm 1}(\theta)+2n\pi$\\[5pt]
        \hline\hline
    &$\mu^{1}_{+\sigma}$\\
    \hline 
        $\mu^{1}_{++}$ & $(\sigma=-1)\&(\phi=n\pi)$\\
        \hline 
        $\mu^{1}_{-\sigma_2}$ &$\theta= 2n\pi- q_{\sigma_2,\sigma}(\phi)~|~\theta= (2n+1)\pi+q_{\sigma_2,\sigma}(\phi)$ \\[5pt]
    \hline \hline
    &$\mu^{1}_{--}$\\
    \hline
   $\mu^{1}_{-+}$ &$\phi=n\pi$ 
\end{tabular}
\end{table}

\subsection{Crossings at $k_{\pm}$ point}
 Here, we will present the derivation of the analytic lines of bands crossing at $\bm{k}=\bm k_{-}=\frac{2\pi}{3a}(\frac{1}{\sqrt{3}},\,1)$.
 Similar expressions can be derived for $\vect k_{+}=\frac{2\pi}{3a}(-\frac{1}{\sqrt{3}},\,1)$.

The easiest way to find the eigenvalues is to diagonalise the nearest-neighbour Hamiltonian, and then diagonalise the next-nearest neighbour one in the degenerate subspaces.
We have two such two-fold degenerate spaces, and the bands are given by the compact form
\begin{equation}\label{eq:eigValsAlpha}
\begin{aligned}
E_{\sigma,0} &=\sigma u,\\
E_{\sigma,u}&=\sigma u-2\sqrt{3} \alpha\sin(\phi),\\
E_{\sigma,v}&=\sigma v+2\sqrt{3} \alpha\sin(\phi),\\
\end{aligned}
\end{equation}
where $u=2\sin(\theta')$ and $v=\sqrt{5+4\cos(2\theta^\prime)}$, where $\theta^\prime=\theta+\pi/3$.
Thus, two energy bands are independent of $\alpha$, and all band energies are linear in the hopping parameters.
Using Eq.~(\ref{eq:eigValsAlpha}), we can find the analytic expressions for the crossing lines. For compactness, we shall use the following short-hand notations:
\begin{equation}
\begin{aligned}
g_{\sigma_1\sigma_2}(\theta')&=\arcsin(\frac{(\sigma_1-\sigma_2)u}{2 \sqrt{3}\alpha}),\\
h_{\sigma_1\sigma_2}(\theta')&=\arcsin(\frac{\sigma_2  u -\sigma_1 v}{2\sqrt{3}\,\alpha}),\\
l_{\sigma_1\sigma_2}(\theta')&=\arcsin(\frac{\sigma_2 u -\sigma_1 v}{4\sqrt{3}\,\alpha}).\\
\end{aligned}
\end{equation}
The analytical expressions for the crossing lines are presented in Table~II. 
\begin{table}[!htb]\label{table:CrossLnsKm}
    \caption{\label{ct1}The analytical form of the dependence of level crossings at $k_-$ as a function of $\theta$ and $\phi$. For each analytical crossing line, the crossing energy bands are listed. Note that the bands $E_{\pm v}$ do not cross with each other.}
    \begin{tabular}{c|c  }
    \hline
        &$E_{\sigma0}$\\
    \hline
    $E_{+0}$ &$(\sigma=-1)\&(\theta=\frac{2\pi}{3}+\pi \, n)$  \\
    \hline 
    $E_{\sigma_1 u}$&$\phi=2n\pi+g_{\sigma_1 \sigma}(\theta')~|~\phi=(2n+1)\pi-g_{\sigma_1 \sigma}(\theta')$ \\
    \hline 
    $ E_{\sigma_1 v} $&$\phi=2n\pi+h_{\sigma_1 \sigma}(\theta')~|~\phi=(2n+1)\pi-h_{\sigma_1 \sigma}(\theta')$ \\[5pt]
    \hline\hline
        &$E_{\sigma u}$\\
    \hline
   $E_{+ u}$ & $(\sigma=-1)\&(\theta=\frac{2\pi}{3}+\pi \, n)$\\
               \hline
       $ E_{\sigma_1 v} $& $\phi=2n\pi+l_{\sigma_1 \sigma}(\theta')~|~\phi=(2n+1)\pi-l_{\sigma_1 \sigma}(\theta')$ 
    \end{tabular}
\end{table}

In the discussion above, we avoided numbering the energy bands. For example, we used $E_{\pm u}$ instead of $E_{1,\,2}$. This is because the order of the energy bands is unknown.\par
In order to sort the energy bands, one needs to find their numerical values at a given point~$(\theta,\,\phi)$ in the phase diagram. This is how the two band indices, which are associated with each analytic crossing line, were identified in Fig.~\ref{fig:Ct2byt1is4by10}(c).  

\section{Topological phase diagrams for other $t_2/t_1$ ratios}\label{sec:AppendixC}
\begin{figure*}[!htb]
\includegraphics[width=0.7\linewidth]{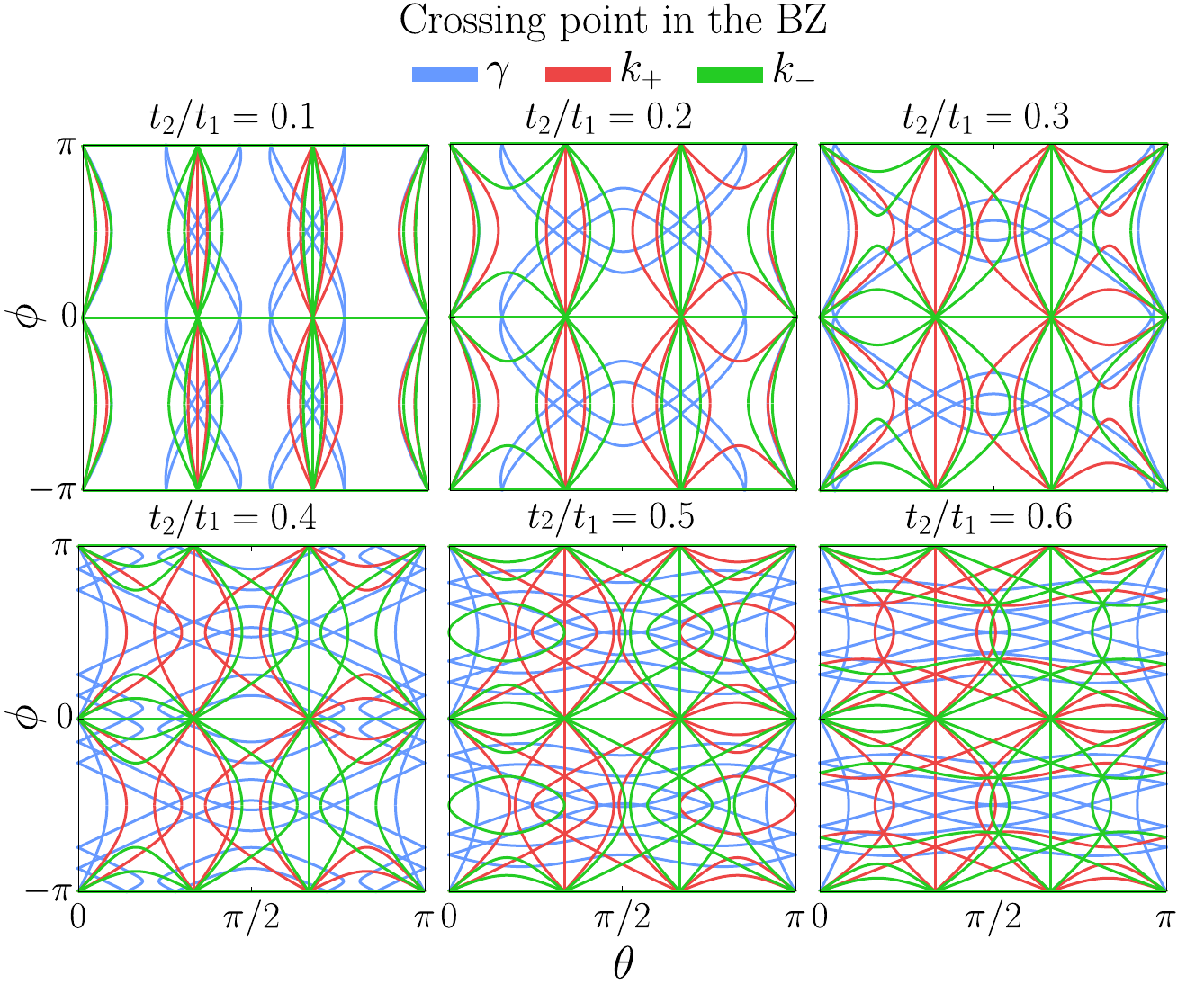}
    \caption{Lines in the topological phase diagram of the generalised Haldane model where \textit{any} two energy bands cross at a high symmetry point in the FBZ, i.e., at $\bm{k}=\gamma,\,k_{+}$ or $k_{-}$. The lines are plotted for different $t_2/t_1$ ratios. }
    \label{fig:AnaCrossLnsDifft2byt1}
\end{figure*}

Fig.~\ref{fig:AnaCrossLnsDifft2byt1} shows the analytical crossing lines for different $t_2/t_1$ ratios. As can be seen in Fig.~\ref{fig:AnaCrossLnsDifft2byt1}, as $t_2/t_1$ increases, the number of band crossings increases; and hence, the phase diagrams become more complex. We used $t_2/t_1=4/10$ for calculating the topological phase diagram in the main text~(see Fig.~\ref{fig:Ct2byt1is4by10}) as this ratio shows many band crossings. The large number of crossings is useful for comparing the topological phase diagrams with the results of the analytical crossing lines.

\begin{figure*}[!htb]
\includegraphics[width=0.7\linewidth]{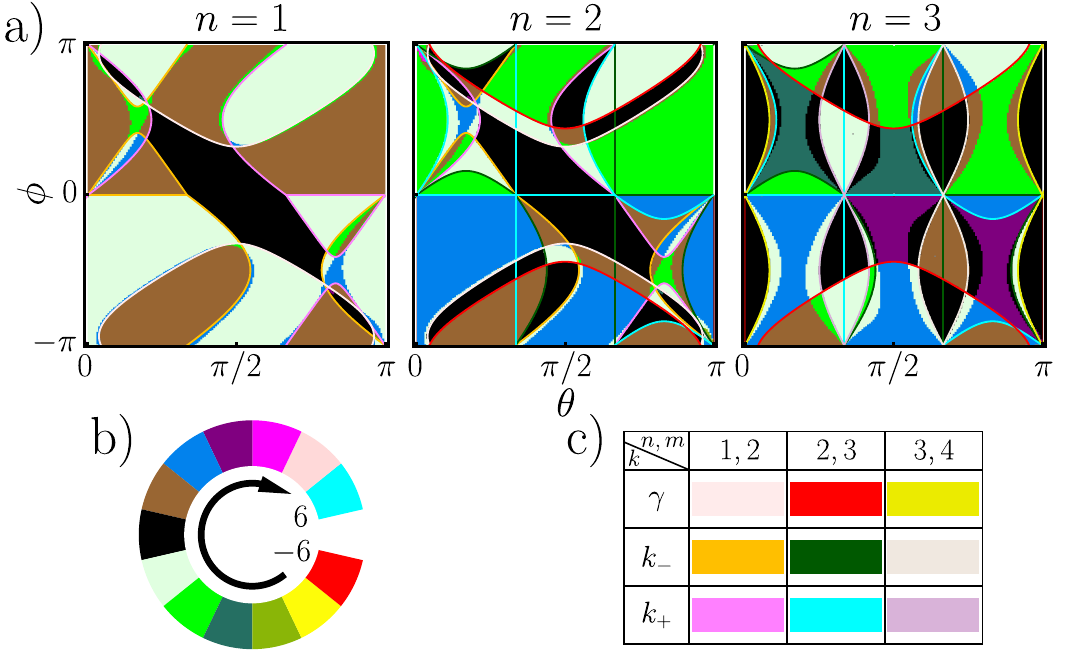}
    \caption{(a) The topological phase diagrams for the first three bands of the generalised Haldane model for $t_2/t_1=3/10$. The details of the calculations are presented in the caption of Fig.~\ref{fig:Ct2byt1is4by10}. (b) The colour bar of the Chern numbers. (c) A table of the analytic crossing lines, showing the crossing point in the BZ and the band indices of the crossing bands.}
    \label{fig:Ct2byt1is3by10}
\end{figure*}

Although the topological phase diagrams for ratios higher than $t_2/t_1=4/10$ are not studied in this work, such diagrams appear to be at least as complex as the topological phase diagram for $t_2/t_1=4/10$, as can be concluded from the number of crossings in Fig.~\ref{fig:AnaCrossLnsDifft2byt1}. The topological phase diagrams for the first three bands, which are results of Fig.~\ref{fig:Ct2byt1is4by10}, are reproduced in Fig.~\ref{fig:Ct2byt1is3by10} for $t_2/t_1=3/10$. As can be seen in Fig.~\ref{fig:Ct2byt1is3by10}, for $t_2/t_1=3/10$, the topological phase diagrams are much simpler than those for $t_2/t_1=4/10$.

\section{The transformation of the Chern number under $(\theta,\,\phi)\rightarrow-(\theta,\,\phi)$}\label{sec:AppendixD}
Under $(\theta,\,\phi)\rightarrow-(\theta,\,\phi)$, the Hamiltonian transforms as $H_{\vect{k}}(\theta,\,\phi)\rightarrow H_{\vect{k}}(-\theta,\,-\phi)=H^*_{-\vect{k}}(\theta,\,\phi)$. Additionally, under such a transformation, the Berry curvature in Eq.~(\ref{eq:BerryCurvature})) transforms as
\begin{widetext}
\begin{equation}
\begin{aligned}
&\Omega_n(\vect{k},-\theta,-\phi) =-2 \Im \sum_{n^\prime\neq n} \frac{\matrixel{u_{n\vect{k}}(\theta,\,\phi)}{\partial_x H_{\vect{k}}(\theta,\,\phi)}{u_{n^\prime\vect{k}}(\theta,\,\phi)}\matrixel{u_{n^\prime\vect{k}}(\theta,\,\phi)}{\partial_y H_{\vect{k}}(\theta,\,\phi)}{u_{n\vect{k}}(\theta,\,\phi)}}{(E_{n\vect{k}}(\theta,\,\phi)-E_{n^\prime\vect{k}}(\theta,\,\phi))^2}\rightarrow\\
&=-2 \Im \sum_{n^\prime\neq n} \frac{\matrixel{u_{n\vect{k}}(-\theta,\,-\phi)}{\partial_x H_{\vect{k}}(-\theta,\,-\phi)}{u_{n^\prime\vect{k}}(-\theta,\,-\phi)}\matrixel{u_{n^\prime\vect{k}}(-\theta,\,-\phi)}{\partial_y H_{\vect{k}}(-\theta,\,-\phi)}{u_{n\vect{k}}(-\theta,\,-\phi)}}{(E_{n\vect{k}}(-\theta,\,-\phi)-E_{n^\prime\vect{k}}(-\theta,\,-\phi))^2}\\
&=-2 \Im \sum_{n^\prime\neq n} \frac{\matrixel{u^*_{n-\vect{k}}(\theta,\,\phi)}{\partial_x H_{-\vect{k}}^*(\theta,\,\phi)}{u^*_{n^\prime-\vect{k}}(\theta,\,\phi)}\matrixel{u^*_{n^\prime-\vect{k}}(\theta,\,\phi)}{\partial_y H_{-\vect{k}}^*(\theta,\,\phi)}{u^*_{n-\vect{k}}(\theta,\,\phi)}}{(E_{n-\vect{k}}(\theta,\,\phi)-E_{n^\prime-\vect{k}}(\theta,\,\phi))^2}\\
&=-2 \Im \sum_{n^\prime\neq n} \frac{\matrixel{u_{n-\vect{k}}(\theta,\,\phi)}{\partial_x H_{-\vect{k}}(\theta,\,\phi)}{u_{n^\prime-\vect{k}}(\theta,\,\phi)}^*\matrixel{u_{n^\prime-\vect{k}}(\theta,\,\phi)}{\partial_y H_{-\vect{k}}(\theta,\,\phi)}{u_{n-\vect{k}}(\theta,\,\phi)}^*}{(E_{n-\vect{k}}(\theta,\,\phi)-E_{n^\prime-\vect{k}}(\theta,\,\phi))^2}\\
&=-2 \Im \sum_{n^\prime\neq n} \frac{\matrixel{u_{n^\prime-\vect{k}}(\theta,\,\phi)}{\partial_x H_{-\vect{k}}(\theta,\,\phi)}{u_{n-\vect{k}}(\theta,\,\phi)}\matrixel{u_{n-\vect{k}}(\theta,\,\phi)}{\partial_y H_{-\vect{k}}(\theta,\,\phi)}{u_{n^\prime-\vect{k}}(\theta,\,\phi)}}{(E_{n-\vect{k}}(\theta,\,\phi)-E_{n^\prime-\vect{k}}(\theta,\,\phi))^2}\\
&=-\Omega_n(-\vect{k},\,\theta,\,\phi).
\end{aligned}
\end{equation}
\end{widetext}
Since $\int_{\in\text{FBZ}} d\vect{k}\,\Omega_n(\vect{k})=\int_{\in\text{FBZ}} d\vect{k}\,\Omega_n(-\vect{k})$, the Chern number, which is the integral of the Berry curvature over the first Brillouin zone times a constant, acquires a minus sign under the transformation $(\theta,\,\phi)\rightarrow-(\theta,\,\phi)$.

\end{document}